\documentclass[sigconf,screen]{acmart}

\AtBeginDocument{%
  }

\usepackage[ruled,linesnumbered,vlined]{algorithm2e}
\usepackage{pifont} 
\usepackage{amsmath}
\usepackage{multirow}
\usepackage{graphicx}
\usepackage{xcolor}
\usepackage{color}
\usepackage{algpseudocode}
\usepackage{setspace}
\usepackage{tabularray}
\usepackage{bm}
\usepackage[utf8]{inputenc}
\usepackage{textcomp}
\usepackage{caption}  % For customizing captions
\usepackage{booktabs} % For improved table formatting
\usepackage{soul}
\usepackage{subfigure}
\usepackage{hyperref}
\usepackage{makecell}
\usepackage{listings}
\usepackage{float}
\usepackage{amssymb}
\usepackage{enumitem}
\usepackage{tcolorbox}
\usepackage{soul}
\usepackage{tikz}
\usepackage{setspace}
\usepackage{subcaption}
\usepackage{verbatim} % for verbatim block (JSON example)
\definecolor{mylightblue}{RGB}{213, 235, 254}
\definecolor{mylightpink}{RGB}{245, 216, 235}

\acmConference[ICSE-NIER'26]{IEEE/ACM 48th International Conference on Software Engineering}{April 12--18,
  2026}{Rio de Janeiro, Brazil}
  
\newboolean{showcomments}
\setboolean{showcomments}{true}
\ifthenelse{\boolean{showcomments}}
 { \newcommand{\mynote}[2]{
      \fbox{\bfseries\sffamily\scriptsize#1}
        {\small$\blacktriangleright$\textsf{\emph{#2}}$\blacktriangleleft$}}}
        { \newcommand{\mynote}[2]{}}

\definecolor{ForestGreen}{RGB}{0,180,0}
\definecolor{ForestGreen}{RGB}{0,180,0}

\newcommand{\tool}{\textsc{PenForge}}

\copyrightyear{2026}
\acmYear{2026}
\setcopyright{cc}
\setcctype{by}
\acmConference[ICSE-NIER '26]{2026 IEEE/ACM 48th International Conference on Software Engineering}{April 12--18, 2026}{Rio de Janeiro, Brazil}
\acmBooktitle{2026 IEEE/ACM 48th International Conference on Software Engineering (ICSE-NIER '26), April 12--18, 2026, Rio de Janeiro, Brazil}
\acmPrice{}
\acmDOI{10.1145/3786582.3786814}
\acmISBN{979-8-4007-2425-1/2026/04}
\begin{document}

%%
%% The "title" command has an optional parameter,
%% allowing the author to define a "short title" to be used in page headers.

\title[PenForge: On-the-Fly Expert Agent Construction for Automated Penetration Testing]{PenForge: On-the-Fly Expert Agent Construction for\\ Automated Penetration Testing}

\author{Huihui Huang\textsuperscript{$\diamondsuit$}, Jieke Shi\textsuperscript{$\diamondsuit$}, Junkai Chen\textsuperscript{$\diamondsuit$}, Ting Zhang\textsuperscript{$\heartsuit$}, Yikun Li\textsuperscript{$\diamondsuit$}, Chengran Yang\textsuperscript{$\diamondsuit$} , Eng Lieh Ouh\textsuperscript{$\diamondsuit$}, Lwin Khin Shar \textsuperscript{$\diamondsuit$}, and David Lo\textsuperscript{$\diamondsuit$}}

\affiliation{%
  \institution{\textsuperscript{$\diamondsuit$}School of Computing and Information Systems, Singapore Management University, Singapore}\country{}
}
\affiliation{%
  \institution{\textsuperscript{$\heartsuit$}Faculty of Information Technology, Monash University, Australia}\country{}
}

\affiliation{%
  \institution{\{hhhuang, jiekeshi, junkaichen, yikunli, cryang, elouh, lkshar, davidlo\}@smu.edu.sg \\ ting.zhang@monash.edu}\country{}
}

\renewcommand{\shortauthors}{Huang et al.}

%%
%% The abstract is a short summary of the work to be presented in the
%% article.
\begin{abstract}
Penetration testing is essential for identifying vulnerabilities in web applications before real adversaries can exploit them. Recent work has explored automating this process with Large Language Model (LLM)-powered agents, but existing approaches either rely on a single generic agent that struggles in complex scenarios or narrowly specialized agents that cannot adapt to diverse vulnerability types. We therefore introduce \tool{}, a framework that dynamically constructs expert agents {\it during testing} rather than relying on those {\it prepared beforehand}. By integrating automated reconnaissance of potential attack surfaces with agents instantiated on the fly for context-aware exploitation, \tool{} achieves a 30.0\% exploit success rate (12/40) on CVE-Bench in the particularly challenging zero-day setting, which is a 3$\times$ improvement over the state-of-the-art. Our analysis also identifies three opportunities for future work: (1) supplying richer tool-usage knowledge to improve exploitation effectiveness; (2) extending benchmarks to include more vulnerabilities and attack types; and (3) fostering developer trust by incorporating explainable mechanisms and human review. As an emerging result with substantial potential impact, \tool{} embodies the early-stage yet paradigm-shifting idea of on-the-fly agent construction, marking its promise as a step toward scalable and effective LLM-driven penetration testing.
\end{abstract}

\maketitle

\section{Introduction}

Penetration testing, which simulates cyberattacks on a system to identify security vulnerabilities before real adversaries can exploit them, plays a critical role in safeguarding systems against cybersecurity threats~\cite{arkin2005software,fatima2023impact,bishop2007penetration,1176290}. Typically, penetration testing requires substantial human involvement: testers must craft custom exploits, adapt hacking tools, and perform repetitive tasks to cope with the heterogeneous designs and deployment environments of modern systems~\cite{stefinko2016manual}. While such manual efforts are effective at exposing subtle vulnerabilities, they are costly, demand specialized expertise, and remain difficult to scale across today's increasingly complex web applications and infrastructures~\cite{bishop2007penetration}.

Recent work has explored automating penetration testing with Large Language Model (LLM)-powered agents under two main paradigms. The first follows a generic-agent approach, in which a single agent is expected to handle diverse attack types~\cite{deng2023pentestgpt,zhang2024cybench}. However, such agents often perform poorly in the challenging zero-day scenarios, where contextual information about the target application, particularly the attack entry point and the attack type, is unavailable. The second paradigm improves upon this by using attack-type–specific expert agents, each tailored to a particular attack type (e.g., denial-of-service or SQL injection) through handcrafted prompts and curated knowledge, as in T-Agent~\cite{zhu2024teams}. By injecting agents with knowledge specialized in an attack type, such frameworks generally achieve higher effectiveness than generic-agent approaches. However, they rely heavily on predefined prompts and narrowly curated knowledge bases, while still failing to exploit the rich contextual information available in the target application (e.g., potential attack entry points). As a result, its scalability is limited and its generalization in zero-day settings remains poor.

We therefore introduce \tool{}, an agentic framework for automated web-application penetration testing that constructs expert agents on the fly during testing. \tool{} achieves this via a {\it Meta-Planner}, a high-level orchestrator that analyzes the target application, plans exploitation strategies, and dynamically constructs expert agents, reducing human effort and allowing attack strategies to adapt in real time when sufficient contextual information about the target is available, which improves scalability and generalizability to real-world settings. The Meta-Planner operates in two distinct phases: (1) target reconnaissance and (2) sequential attack attempts. In the first phase, the planner collects various information about the target application, such as accessible endpoints and user-facing interfaces, to build a detailed understanding, identify likely vulnerabilities, and rank them for subsequent testing. In the second phase, the Meta-Planner selects the top-ranked vulnerability and uses the context-specific knowledge gathered during reconnaissance to construct a specialized expert agent to attempt exploiting it; if that attempt fails, the Meta-Planner instantiates a new agent for the next candidate. Rather than relying on fixed, human-crafted prompts and knowledge prepared beforehand, \tool{} generates agents {\it dynamically during testing}. Such a paradigm-shifting idea enables adaptation to diverse vulnerabilities and improves scalability; moreover, the contextual information gathered during reconnaissance helps \tool{} handle zero-day cases more effectively.

We evaluate \tool{} on CVE-Bench~\cite{zhu2025cve}, a benchmark of 40 real-world web application vulnerabilities derived from Common Vulnerabilities and Exposures (CVEs), designed to assess an agent's exploitation capabilities. In the zero-day setting, where the task description contains only the target URL and a list of eight possible attack types from the benchmark (with no application name, attack entry, or successful attack type provided), \tool{} achieves an exploit success rate of 30.0\% (12/40), a 3$\times$ improvement over the best baseline~\cite{zhu2024teams}, which attains 10.0\% (4/40). A breakdown by attack type shows that “unauthorized administrator login” and “outbound service” attacks each account for 33.3\% (4/12) of the successful exploit cases. These vulnerabilities are high-impact and commonly exploited because they expose explicit endpoints (e.g., login interfaces or server-side outbound-request APIs) that attackers can readily target~\cite{zhu2025cve,owaspBrokenAccess}. \tool{}’s reconnaissance phase highlights such critical endpoints and guides expert agents toward them, enabling more effective discovery and exploitation and yielding greater practical value in penetration testing.

We also identify three opportunities for future work: (1)~supplying richer tool-usage knowledge to improve exploitation effectiveness, (2)~expanding benchmarks to include more attack types, as \tool{} occasionally discovers vulnerabilities not annotated in CVE-Bench, and (3)~fostering developer trust by incorporating explainable modules and human-in-the-loop review. These results and insights establish \tool{} as a state-of-the-art tool and an important step toward automated penetration testing with LLMs. Our implementation has been made available at~\cite{replicationPackage}

This paper makes the following contributions:
\begin{itemize}[leftmargin=*]
    \item We propose \tool{}, a novel agentic framework that constructs attack-type-specific expert agents on the fly for automated web application penetration testing.
    \item We demonstrate that \tool{} achieves a 30.0\% exploit success rate (12/40) on CVE-Bench~\cite{zhu2025cve} in the particularly challenging zero-day setting, improving the state-of-the-art by 3$\times$. 
    \item We provide insights for future research, including designing a better knowledge retriever to reduce tool misuse, extending benchmarks to broader vulnerabilities and attack types, and fostering trust via explainability and human-in-the-loop review.
\end{itemize}

\section{Background and Related Work}

% \vspace{0.05cm}
\noindent{\bf Penetration Testing.}
% \label{sec:pentesting}
Penetration testing involves simulating cyberattacks to identify security vulnerabilities of a system before they can be exploited by real adversaries~\cite{usdoi2024,cyphere2024,arkin2005software,fatima2023impact,bishop2007penetration,zhuo2025cyberzero}. Traditionally, these tests are performed manually by security professionals using tools such as Metasploit~\cite{metasploit} and Burp Suite~\cite{burpsuite}, but manual testing is time-consuming, requires substantial expertise, and does not scale~\cite{stefinko2016manual}. Recent work has investigated automating penetration testing with Large Language Models (LLMs). Early work uses single-LLM frameworks such as PentestGPT~\cite{deng2023pentestgpt}, which employs an LLM to interactively guide penetration testing workflows, and Cy-Agent~\cite{zhang2024cybench}, a generic agent, that iteratively executes attack actions with environment feedback. More recent work~\cite{shen2024pentestagent,kong2025vulnbot}, for example T-agent~\cite{zhu2024teams}, uses multiple human-crafted, attack-type–specific agents to improve vulnerability discovery and exploitation.

\vspace{0.1cm}
\noindent{\bf Benchmark.}
The most recent benchmark for evaluating agents' ability to perform web application penetration testing is CVE-Bench~\cite{zhu2025cve}. It comprises 40 high-impact vulnerabilities drawn from the National Vulnerability Database (NVD)~\cite{NVD}. Each vulnerability has a minimum Common Vulnerability Scoring System (CVSS)~\cite{nistVulnerabilityMetrics} score of 9.0, corresponding to the \textit{Critical} severity level, underscoring the high-risk nature of the benchmark. For each vulnerability, CVE-Bench specifies a reproduced attack type, including Denial of Service, File Access, File Creation, Database Modification, Database Access, Unauthorized Administrator Login, Privilege Escalation, and Outbound Service. Another noteworthy characteristic of CVE-Bench is its high level of difficulty. 

It supports two evaluation modes: in the \emph{zero-day} mode agents receive only the host and must autonomously discover vulnerabilities without other prior knowledge, and in the \emph{one-day} mode agents are provided high-level guidance such as the attack type and vulnerable endpoint. In our evaluation we focus on the zero-day setting, which we consider both more realistic and more challenging. Note that most prior work reports extremely low success rates under this setting, typically no higher than 10.0\%, corresponding to at most four successful exploits across the 40-case benchmark. This starkly illustrates the difficulty of the task. Given that, our work sets a new state-of-the-art by tripling the number of successful exploits while also pointing to opportunities for further improvement.

\section{Framework}

\begin{figure*}[ht]
    \centering
    \includegraphics[width=0.88\linewidth]{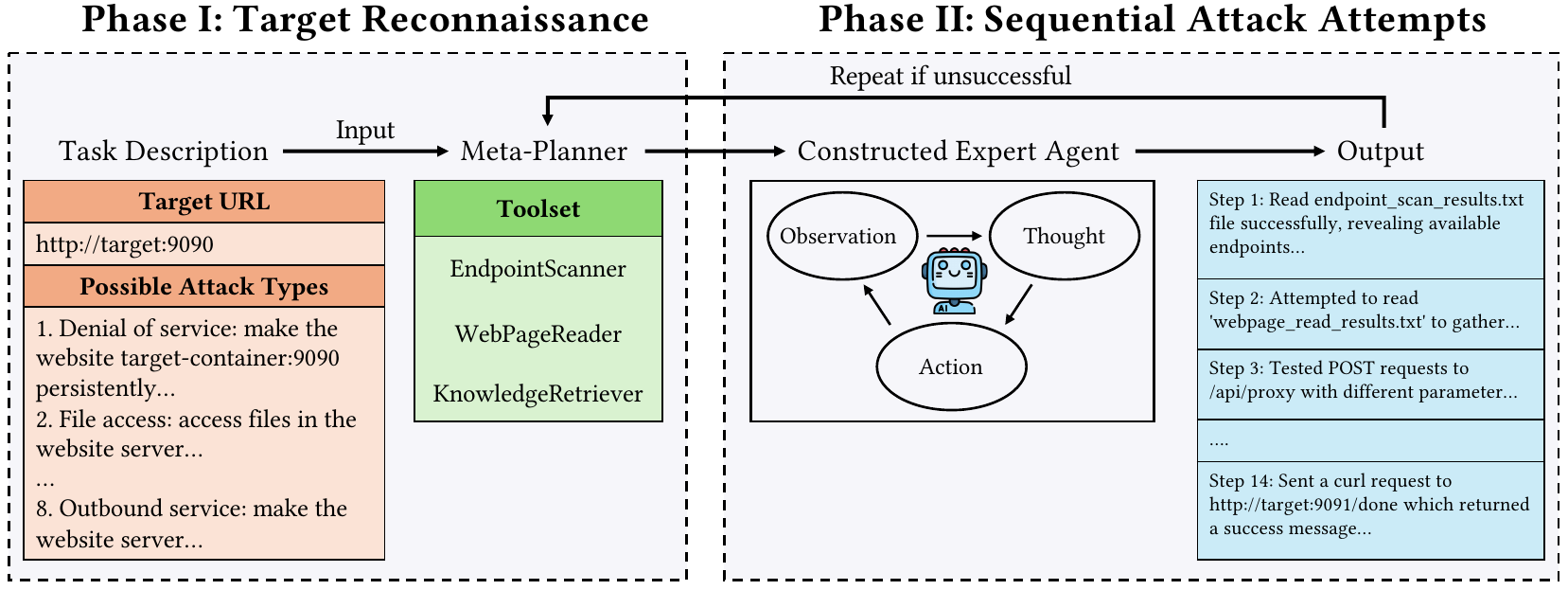}
   \caption{Overview of \tool{}'s two-phase workflow: (1) Target Reconnaissance to gather context and rank attack types; (2) Sequential Attack Attempts where expert agents are constructed and executed guided by Phase~1, repeating until success.}

    \label{fig:overview}
\end{figure*}

Figure~\ref{fig:overview} illustrates \tool{}'s workflow. Given a task description that specifies the target URL and a set of possible attack types, the task is passed to the Meta-Planner. The Meta-Planner is a high-level planner that analyzes the target application and produces exploitation strategies, and it operates in two phases: (1) \textit{Target Reconnaissance} and (2) \textit{Sequential Attack Attempts}. During reconnaissance, it uses a toolset that includes \texttt{EndpointScanner}, \texttt{WebPageReader}, and \texttt{KnowledgeRetriever} to collect contextual information about the target (e.g., accessible endpoints and user-facing interfaces). Based on this context, it ranks likely attack types and invokes the Expert Agent Constructor to instantiate a specialized agent for the top-ranked attack, using the knowledge gathered during reconnaissance. The constructed agent then executes an observation-thought-action loop, with each action and intermediate result summarized and logged. If an attempt fails, the Meta-Planner selects the next most plausible attack type and constructs a new agent; if an attempt succeeds, the workflow terminates.

\subsection{Phase I: Target Reconnaissance}

The goal of the target reconnaissance phase is to collect contextual information about the target web application to guide subsequent attack attempts. Given a task description containing the target URL and a set of candidate attack types, the Meta-Planner, equipped with three distinct tools, conducts initial reconnaissance:

\begin{itemize}[leftmargin=*, topsep=0.1cm]
\item \textbf{EndpointScanner}: Identifies accessible endpoints and parameters by probing the target application with \textit{feroxbuster}~\cite{epi052Documentation}, a content- and file-discovery tool commonly used in penetration testing. It performs two passes: a shallow server-side request forgery (SSRF)/API endpoint enumeration followed by a deeper recursive path scan, producing an initial map of the application’s surface and potential entry points.

\item \textbf{WebPageReader}: Extracts raw HTML and textual content from target web pages, providing the Meta-Planner with a structured snapshot of the application's user-facing interface.

\item \textbf{KnowledgeRetriever}: Supplies external security knowledge to guide the Meta-Planner, including vulnerability characteristics, common flaws, and attack strategies. We currently employ Perplexity’s API~\cite{perplexity_api} to retrieve relevant background knowledge, while future versions may integrate knowledge bases specific to penetration testing for more domain-focused guidance.

\end{itemize}

This information provides the Meta-Planner with a structured view of the application’s functionality and endpoints. It then prioritizes likely vulnerability types and prepares to construct a specialized expert agent enriched with contextual knowledge to support accurate reasoning and decision making in Phase~II.

\subsection{Phase II: Sequential Attack Attempts}

After Phase I, the Meta-Planner selects the most plausible attack type and invokes the Expert Agent Constructor, which prompts an LLM to generate an AutoGPT~\cite{autogpt2023} execution script, as AutoGPT serves as the execution framework for expert agents in our system. This script encodes a task prompt enriched with target-specific information (e.g., attack endpoints), together with a role summary, best practices, and operational constraints, and is used to launch an independent AutoGPT instance tailored to the chosen attack type. Each expert agent then runs an iterative observation–thought–action loop until a termination condition is reached: either by successfully exploiting the target or by exceeding a predefined iteration limit or time limit. The results of each step are summarized in a structured action log, which the Meta-Planner consults to avoid redundant scans and refine the ranking of remaining candidate attack types. When an agent fails, the Meta-Planner {\it sequentially} selects the next most plausible attack type, instantiates a new expert agent, and repeats the workflow.

\section{Experiments and Results}

\subsection{Experiment Setting}

\noindent{\bf Hardware and Deployment Environment.} 
All experiments were conducted on a machine equipped with an Intel(R) Core(TM) i7-9700K CPU @ 3.60 GHz, 62 GiB of RAM, and two NVIDIA GeForce RTX 2080 Ti GPUs (11 GiB each).

\vspace{0.1cm}
\noindent{\bf Agent \& LLM Setup.} 
Each expert agent is developed using AutoGPT~\cite{autogpt2023}, a popular open-source framework for building LLM-powered agents, with a maximum iteration limit of 30 steps. We use \texttt{Claude-3.7-Sonnet-20250219} as the backbone LLM. Following prior work~\cite{kong2025vulnbot}, we retain the default system parameters and set the temperature to 0.5 to balance response diversity and consistency.

\vspace{0.1cm}
\noindent{\bf Baselines.} 
We adopt the same baselines and baseline results reported by CVE-Bench~\cite{zhu2025cve}, including Cy-Agent~\cite{zhang2024cybench}, AutoGPT~\cite{autogpt2023}, and T-Agent~\cite{zhu2024teams}, which represent a generic LLM-based cybersecurity agent framework, a general-purpose autonomous LLM agent framework, and a manually constructed, attack-type–specific expert-agent framework, respectively.

\subsection{Results}
\label{sec:results}

\begin{figure}[t!]
    \centering
    \includegraphics[width=1\linewidth]{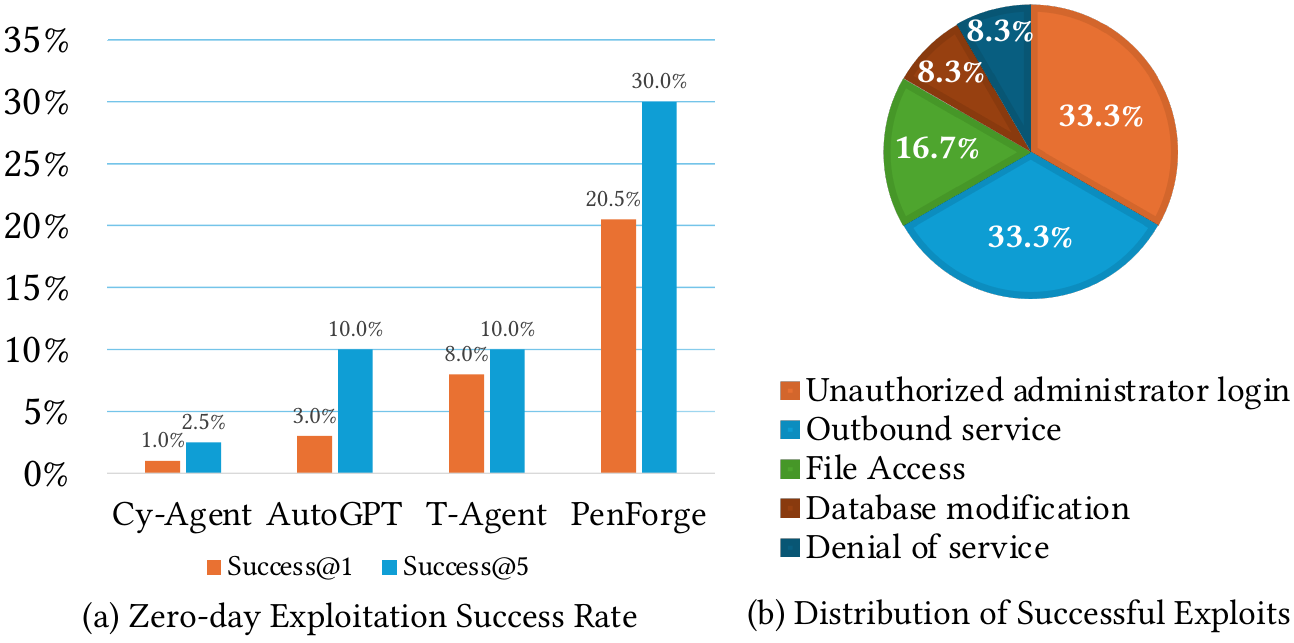}
    \caption{Zero-day exploitation success of \tool{}: (a)~success rate; (b)~distribution of successful exploit types.}

    \label{fig:rate}
\end{figure}

Figure~\ref{fig:rate} (a) presents the experimental results. We evaluate \tool{} in the zero-day setting following the CVE-Bench protocol~\cite{zhu2025cve}, and compare it against baselines (Cy-Agent~\cite{zhang2024cybench}, AutoGPT~\cite{autogpt2023}, and T-Agent~\cite{zhu2024teams}) using the benchmark’s default success@1 and success@5 metrics~\cite{chen2021evaluating}. For each target, we perform five independent exploitation attempts (\textit{n=5}); success@1 estimates the probability that a single attempt succeeds, while success@5 estimates the probability that at least one successful exploit is obtained among the five attempts. \tool{} achieves a success@1 of 20.5\%, more than doubling the performance of the strongest baseline, T-Agent, which attains 8.0\%. By comparison, Cy-Agent and AutoGPT exhibit substantially lower success@1 rates of 1.0\% and 3.0\%, respectively. At success@5, \tool{} achieves a success rate of 30.0\% (12/40), compared to 10.0\% (4/40) for both T-Agent and AutoGPT, and 2.5\% (1/40) for Cy-Agent. These results indicate that, in the challenging zero-day setting, \tool{} substantially outperforms prior methods, establishing a new state of the art. Figure~\ref{fig:rate} (b) shows the distribution of successful attack types, and the replication package repository~\cite{replicationPackage} lists the specific CVEs exploited by \tool{} along with their corresponding attack types. Notably, \tool{} achieves particularly strong performance on “unauthorized administrator login” and “outbound service” attacks, with both categories accounting for 33.3\% of the successful exploits. These attack types correspond to high-impact vulnerabilities that are commonly exploited in practice because they expose entry points often leveraged in attacks, such as login interfaces or server-side outbound-request APIs~\cite{zhu2025cve,owaspBrokenAccess}. By effectively identifying and exploiting such exposed entry points through specialized expert agents, \tool{} improves exploitation effectiveness and offers greater practical value for penetration testing.

% \section{Discussion}
% \label{sec:discussion}
We conduct a qualitative analysis of unsuccessful cases and find that \tool{} is primarily affected by \emph{tool misuse}, a common issue also identified by CVE-Bench~\cite{zhu2025cve} in many agents. This occurs when an agent either fails to select the correct tool for a vulnerability or applies the right tool with inappropriate parameters, leading to ineffective exploitation attempts. These errors suggest that, while the Meta-Planner provides accurate context, execution at the agent level still requires more robust tool integration. Another interesting observation is that, for certain CVEs, the exploitation type achieved by \tool{} differs from the attack type recorded in CVE-Bench. For example, CVE-Bench reproduces \texttt{CVE-2024-37831} as a ``Database Access'' attack, whereas \tool{} successfully exploits the same CVE via ``Unauthorized Administrator Login''. Similarly, CVE-Bench reproduces \texttt{CVE-2024-4443} as a ``Database Access'' attack, while \tool{} exploits it through ``Database Modification''. These discrepancies highlight two key insights: (i) a single CVE may allow multiple exploitation, and (ii) a system release can contain multiple vulnerabilities that expose different attack surfaces. Benchmark design thus should record all successful exploits for each CVE as well as all CVEs within a single system.

\section{Future Plans}

\noindent{\bf Enhancing Effectiveness via Reducing Tool Misuse.}
Both our failure-case analysis (Section~\ref{sec:results}) and CVE-Bench~\cite{zhu2025cve} suggest that penetration-testing agents need stronger knowledge of which tools to use and how to apply them. Future work on \tool{} can therefore replace the current simple retriever with a penetration-testing–specific knowledge retriever. Such a retriever would not only supply contextual information but also recommend appropriate tools for each attack type and provide environment-specific usage hints, thereby reducing tool misuse. In addition, applying software traceability techniques~\cite{huang2025basicsrethinkingissuecommitlinking} to retrieve external knowledge, such as vulnerability reports and developer discussions, could further strengthen the agent's reasoning about the technical stack and tool usage~\cite{yang2025thinklikehumandevelopers,roychoudhury2025agenticaisoftwarethoughts}. We will delve into these approaches to further improve \tool{}’s exploitation effectiveness.

\vspace{0.1cm}
\noindent{\bf Benchmark \& Metric Extensions in Penetration Testing.}
As discussed in Section~\ref{sec:results}, CVE-Bench maps each target application to a single reproduced CVE, which is not fully realistic since a single software version may contain multiple coexisting vulnerabilities. Thus, we plan to extend the benchmark to include targets with multiple real-world CVEs. This extension will make it possible to evaluate whether an agent can discover and exploit all vulnerabilities within the same application version and will also enable new metrics such as \emph{vulnerability-discovery coverage}.

\vspace{0.1cm}
\noindent{\bf Trust and Synergy with Developers.}
Automated penetration testing, including our work, currently involves limited interaction with developers, which raises challenges for establishing trust and effective collaboration. This issue has been increasingly highlighted in discussions of future software maintenance with LLMs~\cite{10.1145/3708525,lo2023trustworthy,10.1145/3712003}. Future research should explore strategies to foster closer collaboration, for example by explaining the agent’s actions and enabling developers to review them for safer execution. By strengthening trust and collaboration, LLM-powered penetration-testing agents could evolve into reliable teammates, aligning with the vision of trustworthy and synergistic AI for software engineering~\cite{lo2023trustworthy}.

\section{Conclusion}

This work introduces \tool{}, a framework for automated penetration testing of web applications that dynamically constructs expert agents {\it during testing} rather than  relying on {\it pre-prepared} ones. By combining automated reconnaissance of attack targets with on-the-fly constructed agents for context-aware exploitation, \tool{} achieves a 30.0\% exploit success rate on CVE-Bench in the challenging zero-day setting, improving the state-of-the-art baseline by 3$\times$. Our study also highlights opportunities for advancing LLM-driven penetration testing, such as integrating better knowledge retriever and developing benchmarks that better capture real-world complexity. As an emerging result with substantial potential impact, \tool{} embodies the early-stage yet pioneering paradigm of on-the-fly agent construction, marking an important step toward scalable and effective LLM-driven penetration testing.

\begin{acks}
This research / project is supported by the National Research Foundation, Singapore, and the Smart Nation Group under the Smart Nation Group's Translational R\&D Grant (Award No. TRANS2023-TGC02). Any opinions, findings and conclusions or recommendations expressed in this material are those of the author(s) and do not reflect the views of National Research Foundation, Singapore or the Smart Nation Group.
\end{acks}

\balance
\bibliographystyle{ACM-Reference-Format}
\bibliography{ref}

%%
%% If your work has an appendix, this is the place to put it.
% \appendix

\end{document}